\documentstyle[12pt]{article}
\input{epsf}
\renewcommand{\thesubsection}{\arabic{subsection}}

\newlength{\extraspace}
\newlength{\extraspaces}
\newcounter{dummy}

\newcommand{\be}{\begin{equation}
\addtolength{\abovedisplayskip}{\extraspaces}
\addtolength{\belowdisplayskip}{\extraspaces}
\addtolength{\abovedisplayshortskip}{\extraspace}
\addtolength{\belowdisplayshortskip}{\extraspace}}
\newcommand{\ee}{\end{equation}}

\newcommand{\ba}{\begin{eqnarray}
\addtolength{\abovedisplayskip}{\extraspaces}
\addtolength{\belowdisplayskip}{\extraspaces}
\addtolength{\abovedisplayshortskip}{\extraspace}
\addtolength{\belowdisplayshortskip}{\extraspace}}
\newcommand{\ea}{\end{eqnarray}}

\newcommand{\newsubsection}[1]{
\vspace{13mm}
\pagebreak[3]

\addtocounter{subsection}{1}
\addcontentsline{toc}{subsection}{\protect
\numberline{\arabic{subsection}}{#1}}
\noindent{\sc \thesubsection.
#1}
\nopagebreak
\vspace{2mm}
\nopagebreak}

\def \is {\! & = & \!}
\def \half {{\textstyle{1\over 2}}}
\newcommand{\figuur}[3]{
\begin{figure}[t]\begin{center}
\leavevmode\hbox{\epsfxsize=#2 \epsffile{#1.eps}}\\[3mm]
\parbox{15.5cm}{\small
\it #3}
\end{center}
\end{figure}}

\begin{document}

\newcommand{\QQ}{{\mbox{\small $Q$}}}
\newcommand{\MM}{{\mbox{\small $M$}}}
\newcommand{\PP}{{\mbox{\small $P$}}}

\newcommand{\del}{\partial}
\def \nonu {\nonumber \\[2.5mm]}

\def \ie {{\it i.e.}}
\def \W {{\cal W}}

%***********************************************************

\addtolength{\baselineskip}{.65mm}
\thispagestyle{empty}

\begin{flushright}
{\sc PUPT-2143}\\
hep-th/0411100 \\
November 2004
\end{flushright}

\bigskip
\bigskip
\bigskip

\begin{center}
{\Large{Hawking Effect in 2-D String Theory}}\\[15mm]
{\sc Joshua J. Friess and Herman Verlinde}\\[9mm]
{\it Joseph Henry Laboratories\\[4mm]
 Princeton University, Princeton, NJ 08544}
\\[25mm]

{\sc Abstract}
\end{center}

\noindent
We use the matrix model to study the final state resulting from a coherent high 
energy pulse in 2-d string theory at large string coupling. We show that 
the outgoing signal produced via reflection off the potential has a thermal spectrum, 
with the correct temperature and profile to be identified with Hawking radiation. 
We confirm its origin as geometrical radiation produced by the gravitational background.
%generated by the classical 
%space-time geometry produced by the incoming matter pulse. 
However, for a total incoming energy $M$, the amount of energy carried by the thermal radiation 
scales only as $\log M$.
Most of the incoming energy is returned 
via the transmitted wave, which does not have a thermal spectrum, indicating the absence 
of macroscopic black hole formation. 

\vfill

\def \BB {{\mbox {\small B}}}

\newpage
\newsubsection{Introduction}

The correspondence
between 2-d string theory and the $c=1$ matrix model is the oldest example of a holographic duality
\cite{Polchinski:1994mb}. 
Recently, it has been further clarified via the identification of the matrix model
with the world-line degrees of freedom of unstable D-particles 
\cite{reloaded,Martinec:2003ka,Klebanov:2003km,McGreevy:2003ep,Nakayama:2004vk,Martinec:2004td}.
Given that the classical low energy target space theory supports a 2-d black hole 
space-time \cite{wittenbh,Mandal:1991tz}, it is natural to ask whether the $c=1$ 
S-matrix reveals characteristic features that
would underscore its possible interpretation in terms of black hole formation and evaporation.

The cleanest formulation of the matrix model is in terms of free fermions moving in an 
upside-down harmonic oscillator potential. In this language, the matrix model is exactly soluble 
and admits an explicit S-matrix description \cite{mpr}.  The collective 
perturbations of the Fermi surface are described by a scalar field $\varphi$,
which in turn are related to the space-time profile of the closed 
string fields via a non-local leg-pole transformation. In the case of type 0B non-critical 
string theory, the closed string fields are the tachyon and axion. 
The presence of two space-time scalar fields connects with the fact that the Fermi sea
of the matrix model splits in two halves, one on each side of the inverted harmonic potential
 \cite{newhat,Takayanagi:2003sm}.
For the case of the axion, the leg-pole transform reads
\be
\label{tachdef}
C(u) \, = \, {\Gamma({1\over 2}(1-\, \partial_u)) \over \Gamma({1\over 2}(1+ \partial_u) )} \;
\varphi(u)
%\, , \qquad \qquad
%C_{out}(v) \, = \, {\Gamma(\; \partial_v \; ) \over \Gamma(-\, \partial_v )} \;
%\varphi_{out}(v)\, .
\ee
where $\varphi(u)$ denotes the anti-symmetric combination of $\varphi$ 
\cite{newhat,Takayanagi:2003sm}. This leg pole factor is an essential ingredient in the 
holographic dictionary of the matrix model; in particular, it ensures a causal description of
gravitational bulk scattering \cite{joepn, joepi}.

In this paper, we will study the final state produced by a large incoming
matter pulse. For the incoming state, we take a general coherent state of the
$\varphi$-field
\be
\label{vef}
|\, {\it in}\, \rangle \, = \, V_f \; | \, 0 \, \rangle \, , \qquad \qquad 
V_f = \exp\Bigl(\int\!\! d u\, f(u) \partial_u \varphi(u)\Bigr) 
\ee
Since the leg-pole transformation is linear, this also represents a coherent
state in terms of the space-time fields.  However, a sharply localized incoming matter pulse in 
the target space
amounts to a smeared distribution in terms of the $\varphi$-field.

The calculation of the final
state is quite straightforward \cite{mpr}. First one re-expresses the above
state in terms of the fermion field via the replacement $\partial \varphi 
= \psi^\dagger \psi$. Next one applies the single particle $S$-matrix to each
fermionic mode to obtain the corresponding fermionic {\it out}-state.  
The bosonic target space {\it out}-state is then obtained by rebosonization and 
applying the leg-pole transform. 

\figuur{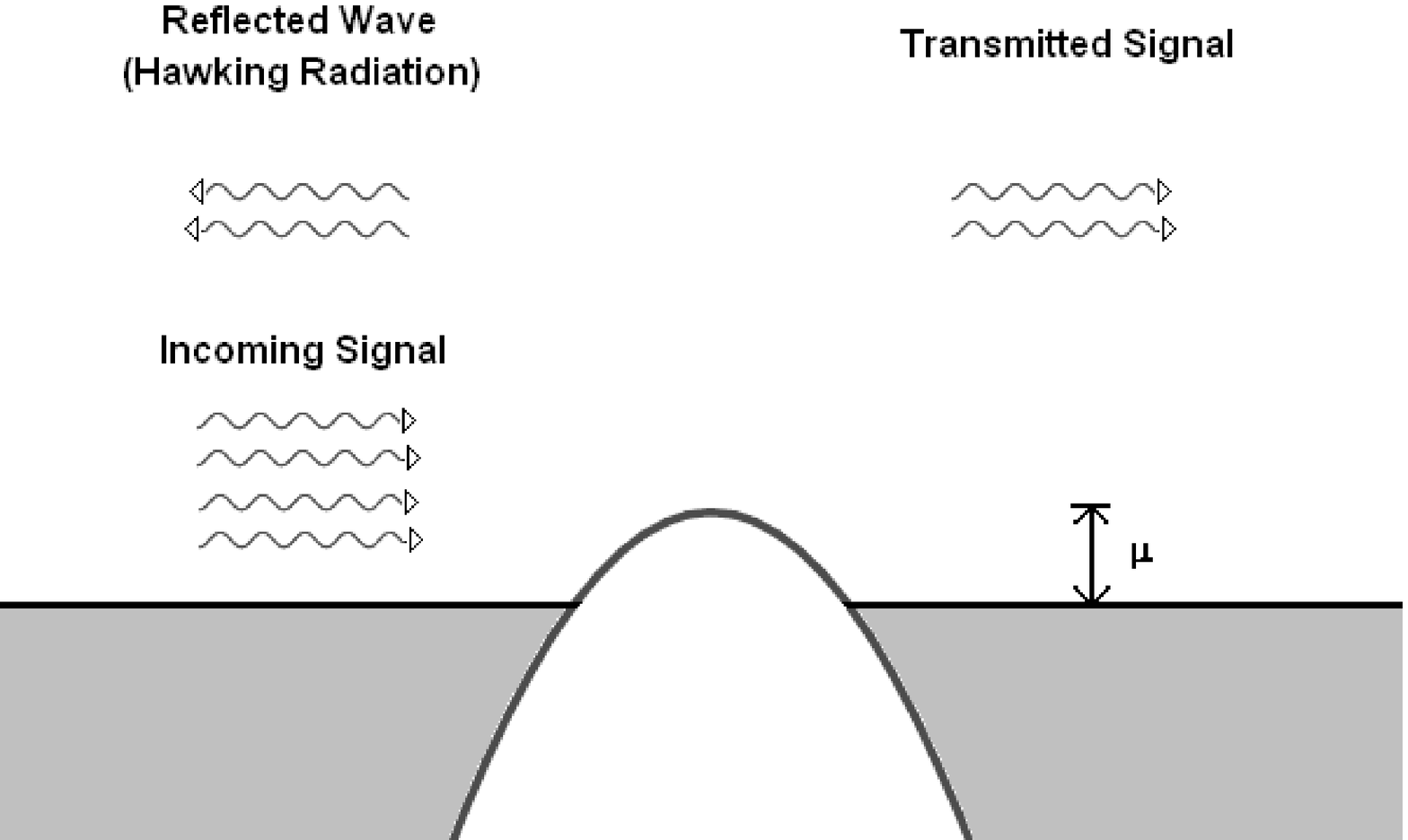}{9cm}{Fig 1: Pictorial description of the scattering process.  Matter 
comes in from one side of the well.  The portion reflecting back has the form of 
Hawking radiation.  Since both asymptotic regions are identified in the target space, 
both the Hawking radiation and the transmitted signal return to the same physical 
spacetime location.}

Our main result will be that the final state contains a clearly identifiable 
thermal component, with the correct temperature to support its interpretation as
Hawking radiation \cite{Hawking:1974sw}.  As indicated in Fig 1, we will find that the thermal 
component consists 
of the matrix model fermion modes that have reflected back off the potential. This reflected
subsector is correlated with the transmitted subsector, and thus forms a mixed state.      
The last two steps of the scattering recipe 
will not alter the thermal nature of this part of the final state: bosonization preserves
thermality, and since the leg-pole factor is just an energy dependent phase, it drops out
of a thermal density matrix.

% (\ref{tachdef}).

This paper is organized as follows.
After reviewing the mechanics of matrix model scattering, we give a short
description of gravitational backreaction and the Hawking effect in 2-d dilaton
gravity. We then show
that the {\it out}-state resulting from a localized
incoming matter pulse contains Hawking radiation.  We will find, however, that the
Hawking radiation carries only a small fraction of the total energy. Most of the
energy is returned via the transmitted signal, which does not have a thermal spectrum.
We give a possible space-time interpretation of the total scattering process in the final section.

Some recent related work on high energy scattering 
and particle creation in 2-d string theory
includes \cite{Karczmarek:2004yc,Das:2004hw,Mukhopadhyay:2004ff,Martinec:2004qt,kms}.

\newsubsection{Scattering as Fourier Transform}

We begin with a description of the exact scattering matrix of the matrix
model given in \cite{mpr}, and highlight its interpretation as a simple
Fourier transformation.

After reduction to the eigenvalues, the $c=1$ matrix model becomes a system of non-interacting
fermions, each described by a single particle Hamiltonian  
\be
\label{ham}
H(p,x) \, =  \half \, p^2
- \half \, x^2, 
\ee
where $x$ denotes the matrix eigenvalue, and $p$ its canonical momentum.
%Here we have chosen the potential $V(x)$ with a small extra quartic term added to
%ensure that it is bounded from below.
The number of fermions equals the rank $N$ of the hermitian matrix. 
To characterize the large $N$ dynamics, it is convenient to
introduce a second quantized fermion language. The fermionic field operators
are expanded as
\ba
\label{expan}
\psi_\alpha(x) \is \int \! {d\omega\over 2\pi} \; b_\alpha(\omega)\; |x|^{-i\omega -\half}\, , \\[2mm]
\psi^\dagger_\alpha
(x) \is  \int \! {d\omega\over 2\pi}
\;  b^\dagger_\alpha(\omega) \; |x|^{i\omega -\half} \nonumber
\ea
where the subscript $\alpha=\pm$ denotes the restriction of $x$ to the positive or negative 
half-line. The modes $b^\dagger_\alpha(\omega)$ and $b_\alpha(\omega)$ respectively create and annihilate a 
quantum mechanical eigenvalue, with a wave function $|x|^{i\omega}$, supported 
on the corresponding half-line. They satisfy the canonical anti-commutation relation
\be
\{ b_\alpha(\omega), b_\beta^\dagger (\xi)\} = 2\pi \delta_{\alpha\beta} \, \delta(\omega-\xi).
\ee
In the ground state of the matrix model, the fermions fill a Fermi sea of states with energy
smaller than the Fermi energy $\mu$:
\ba
b_\alpha(\omega) | \mu \rangle \is 0\, \qquad  \omega > -\mu \, , \\
b^\dagger_\alpha(\omega) | \mu \rangle \is 0\, \qquad  \omega < -\mu \nonumber
\ea
Note that the fermion Hilbert space is similar to that of a field theory in Rindler space. 
This correspondence will play an important role in what follows.

The fermions evolve individually via the Hamiltonian (\ref{ham}).
This Hamiltonian evolution maps asymptotic single particle $in$-states to asymptotic
single particle $out$ states. The resulting single particle $S$-matrix preserves 
energy, and for given $\omega$, acts as a two-by-two matrix \cite{mpr}
\be
\label{smatrix}
\psi^{out}_\alpha(\omega)\; = \; S_\alpha^\beta(\omega)
\, \psi^{in}_\beta(\omega)
\ee
where
\ba
\label{somega} S_{\pm}^{\mp}(\omega) &=&
\frac{1}{\sqrt{2\pi}}\, \mu^{-i\omega}\,
e^{\frac{\pi}{2}\omega}\,
\Gamma(\half-i\omega)\ ,
\nonumber\\[2mm]
S^{\pm}_{\pm}(\omega) &=&
\frac{i}{\sqrt{2\pi}}\,\mu^{-i\omega}\,
e^{-\frac{\pi}{2}\omega}\,
\Gamma(\half-i\omega)\ .
\ea
The relation
\be
|S^\pm_-(\omega)|^2 + |S^\pm_+(\omega)|^2 =1
\ee
shows that the transition from {\it in}\ to {\it out}\ states
preserves probability.

This result for the $c=1$ S-matrix has a simple interpretation as a Fourier transformation.
Let us denote the $in$ coordinate and momentum by $(x,p)$ and the $out$ coordinate and momentum 
by $(y,q)$. We claim that the above $S$-matrix in effect amounts to the identification
$y=p$ and $x=q$, so that the $in$ and $out$ Kruskal coordinates are each others canonical
momenta: $[x,y]=i$. Explicitly, let us introduce two independent $in$ and $out$ one particle
wave functions of frequency $\omega$ via
\ba
\label{vacua}
&& \langle\, x\, |\omega,in \rangle_\pm =
%\frac{1}{\sqrt{2 \pi}} 
\, |x|^{-\half-i\omega}\, \theta(\pm x) \ ,
\nonumber\\[2mm]
&& \langle\, y\, |\omega,out \rangle_\pm = %\frac{1}{\sqrt{2 \pi}} 
\,
|y|^{-\half+i\omega}\, \theta(\pm y) \ . \ea 
%These {\it in}\ and {\it
%out}\ vacua all satisfy ${1\over 2} (xy+yx)|\, 0 \, \rangle = \mu
%\, | \, 0\, \rangle$, and represent the filled Fermi sea. 
These $in$
and $out$ states are linearly dependent. The transition from $x$ to
$y$ is just a Fourier transformation, and it is easily checked
that this leads to the following relation 
\be \label{expec}
{}_\beta\langle \, \omega' ,out 
|\omega ,in \, \rangle_\alpha = S_\alpha^\beta(\omega) \, \delta(\omega - \omega') 
\ee
where $\alpha,\beta=\pm$. 
This representation of the $S$-matrix as a Fourier transform can be directly understood from
its definition as the evolution of a particle in an (inverted) harmonic potential: for a 
right-side
up harmonic oscillator, the wave function evolves into its Fourier transform every quarter of a 
period.
The hyperbolic phase space trajectory that connects the $in$ and $out$ state for the upside-down
harmonic oscillator can be thought of as obtained via analytic continuation from $1/4$
of an elliptic orbit of the right-side up oscillator.

The time-evolution and $S$-matrix preserve an infinite set of conserved quantities
\be
\label{www}
{\cal W}_{mn}
%_{\alpha\beta} 
= \int\limits_{-\infty}^\infty \!\! dx :\! \psi^\dagger (x) \{p^m, x^n\} \psi(x)\!:
\ee
that span a $\W_\infty$-algebra. The S-matrix acts on these charges by exchanging the two labels $n$ and $m$
\be
{\cal W}_{nm}^{out} = 
%S^\dagger_{\alpha\gamma}(H) (
{\cal W}_{mn}^{in}%)_{\gamma\delta} S_{\delta\beta}(H)
\ee
A special case of this equation is conservation of total energy $H = \W_{11}$.
Two charges that will play a special role in the following are 
\ba
\label{kmom}
P \; \equiv\; {\cal W}_{10}  \is  \int\!\!dx :\! \psi^\dagger (x) \, p \, \psi(x)\!  : \nonumber \\[2.5mm]
Q \; \equiv \; {\cal W}_{01} \is   \int\!\!dx :
\! \psi^\dagger (x) \, x \, \psi(x) \! :.
\ea
The operator $P$ is the total incoming Kruskal momentum, and $Q$ is
identified with the total outgoing Kruskal momentum via the S-matrix. As part of the $\W$-algebra, 
they satisfy the ``canonical'' commutation relation
\be
\label{cancan}
[ \, P \, , \, Q \, ] = \Lambda
\ee
with  
\be
\label{lll}
\Lambda \; \equiv \, {\cal W}_{00}  = \,  \int\!dx :\!\psi^\dagger (x) \psi(x)\!  : \;
\ee
the central element of the $\W_\infty$-algebra.
This relation will be of use later on.

\newsubsection{Backreaction and Hawking Effect in 2-d Dilaton Gravity}

Our goal is to compare the $c=1$ matrix quantum mechanics with
the expected gravitational physics of the target space string theory.
By now, there is a relatively
complete understanding of how the weak coupling dynamics of 1+1-d
string theory is encoded in the matrix model \cite{Martinec:2004td,Ginsparg:1993is}.
In the strong
coupling regime, on the other hand, the dictionary is much
less clear. In this regime, %the string coupling $g_s = e^{2\phi}$ becomes
%large much before in coming waves can reflect off the tachyon wall, and
it is expected that gravitational interactions become important.

The gravitational part of the 2-d low energy effective field theory is
described by the well-known dilaton gravity action
\be
\label{act}
S_{grav} =  \int \!
d^2 x \sqrt{-g} e^{-2\phi} \Bigl(R +4 (\nabla \phi)^2 + {\lambda}\Bigl)
%- (\nabla {\cal T})^2 + {4\over \alpha'}\,  {\cal T}^2 \Bigl].
\ee
Here $\lambda = 8/\alpha'$ and $\phi$ is the dilaton. 
This gravitational sector does not support any local dynamics; its equations of motion
are solved by
\ba
\label{lind}
ds^2 \is e^{2\phi} dx^+ dx_-
\\[2.5mm]
%\, , \qquad \qquad
e^{-2\phi} \is  -\lambda x^+ x^- + a\, x^+ + b \, x^- + c\, , \ea
with $a$, $b$ and $c$ integration constants. The vacuum solution
is the linear dilaton background \be \label{lindil} \qquad \quad
e^{-2\phi} =-\lambda x^+x^- \qquad \qquad \pm x^\pm
> 0\, . \ee The coordinates $(x^+,x^-)$ are similar to Kruskal
coordinates, related to the Minkowski light-cone coordinates $(u,v)$ via 
\be
\label{minkdef} (x^+,x^-) = (e^u,-e^{-v})\, . 
\ee 
The string theory dynamics in the linear dilaton background are believed to be
accurately described by the matrix model. 
For general values of the integration constants in (\ref{lind}), however,
the geometry is that of a two-dimensional black hole \cite{wittenbh}.

The central question we wish to address is whether, starting from the matrix model, 
we can recognize this black hole background as the intermediate state produced by 
a high energy incoming matter pulse. In particular, we wish to identify
the component of the final state that describes the Hawking radiation generated
by the black hole geometry. 
This Hawking effect is expected to occur regardless of whether the system actually 
develops a true event horizon or not, except that in case it does not, the thermal
radiation is produced only for a rather short amount of time. Conversely, intermediate 
black hole formation is signalled by a final state that looks thermal for a 
sufficiently long time interval.

Let us recall how the Hawking effect arises from the
perspective of the low energy effective field theory.  
Besides the dilaton gravity fields, 2-d type 0B string theory contains matter 
degrees of freedom in the form of the NS-NS tachyon field ${\cal T}$ and 
the RR axion $C$ \cite{newhat}. To leading order in $\alpha'$, the action for the 
axion is simply that of a massless scalar field
\be
S_C = \int\! d^2 x \sqrt{-g} (\nabla C)^2.
\ee 
Small excitations of the $C$-field are therefore expected to travel along
light-like trajectories, at least in the region of weak curvature and string coupling.

Now consider a large incoming matter pulse traveling along a light-like
trajectory $x^+ = q^+$, with Kruskal momentum $p_+$ and energy
$\MM$. The energy and Kruskal momentum are related via \be \MM =
q^+ p_+ \ee The corresponding space-time geometry is obtained by gluing together, 
along the matter trajectory, the linear dilaton vacuum solution $x^+<
q^+$, to a static black hole solution for $x^+> q^+$ \be
\label{bhv} \, e^{-2\phi} \, = \, -\lambda x^+(x^- +
{1\over\lambda}\, p_+) + M\, . \ee 
Note that for this solution, the dilaton and metric are
continuous along the matter pulse. Assuming that the solution is sufficiently
accurate, it describes the formation of a black hole with a horizon located 
at $x^- = - {1\over \lambda} \, p_+$. It is useful, however, to redefine
coordinates such that the relationship (\ref{minkdef}) between
$x^-$ and the Minkowski light-cone coordinate $v$ is restored, in other words,
to shift the definition of $x^-$ such that the observable part of space-time 
is still given by the region $x^-<0$. In this new coordinate system, right-moving 
geodesics undergo a shift
\be
\label{shifto}
\qquad x^- \rightarrow x^- +\delta x^- \qquad \qquad
\delta x^- = {1\over \lambda}\, p_+
\ee
upon crossing the matter trajectory. As emphasized in particular by
't Hooft and in \cite{svv,kvv}, this shift leads to a characteristic 
commutation relation between the $in$ and $out$-modes.

Let us introduce the $in$ quantum operator $V_{p_+}$ that creates
the incoming matter pulse. We imagine that $V_{p_+}| \, 0 \, \rangle$
is an optimal approximation of an eigenstate of the Kruskal
momentum operator $\hat{\PP}_+$. Now let $\varphi_{out}(x^-)$ denote a
local $out$ operator. If we assume that the shift (\ref{shifto})
is the only direct interaction between the two modes, 
they satisfy an exchange algebra of the form
\be
\label{shft}
\varphi_{{out}}(x^-) \; \hat  {V}_{p_+} \, =\;
 \hat V_{p_+} \; \varphi_{out}(x^-\!\! -\! \textstyle{ 1\over \lambda}\, p_+).
\ee
This relation directly gives rise to the Hawking effect \cite{svv,Giddings:1992ff}.
In terms of wave modes with given frequency
$e^{i\omega v}$, the exchange algebra reads \be \label{exch} a_\omega\,
\hat V_{p_+} = \, \hat  V_{p_+} \; \int\limits_0^\infty {d\xi\over 2\pi}
\,\Bigl( {\alpha}_{\omega\xi}\; a_\xi \, +\,
{\beta}_{\omega\xi} \; a^\dagger_\xi\, \Bigr) \ee\ with \be
\label{bcoef} \beta_{\omega\xi}(p_+) = 
\Bigl({p_+\over\lambda}\Bigr)^{i(\omega+\xi)}\, \sqrt{\xi\over
\omega}\;
{\Gamma(1-i\omega)\Gamma(i(\omega+\xi))\over\Gamma(1+i\xi)} \ee
and $\alpha_{\omega\xi} = \beta_{\omega,-\xi}$. Now let us imagine that $\varphi$ represents
a massless scalar field. The linear combination of creation and annihilation modes on
the right-hand-side then represents a Bogolubov transformation. The
resulting expectation value of the particle number operator is \be
\label{integral}
\langle N(\omega)\rangle \; =
\; \langle \, 0\, |\hat V^\dagger_{p_+}
a^\dagger_\omega a_\omega \hat  V_{p_+}|\, 0\, \rangle \nonumber \; = \;
\int\limits_0^\infty\!{d\xi\over 2\pi} \,|\beta_{\omega\xi}|^2 \, .
\ee
% = \; {1\over 4\pi} \, \int\limits_0^{\infty}\!\!
%{d\xi\over \omega\!+\!\xi} \,
%{\sinh \pi\xi\over \sinh \pi (\omega+\xi)
%\sinh\pi \omega} \, \nonumber \\
%\, \is {1\over 4\pi} \, \int\limits_\omega^{\infty}\!\!
%{d\xi\over \xi} \,
%{\sinh \pi(\xi - \omega)\over \sinh \pi\xi
%\sinh\pi \omega} \, \nonumber \\
%\, \is {1\over 4\pi} \, \int\limits_\omega^{\infty}\!\!
%{d\xi\over \xi} \,
%{{\sinh \pi\xi \cosh \pi\omega - \cosh \pi\xi \sinh \pi\omega} \over \sinh \pi\xi
%\sinh\pi \omega} \, \nonumber \\
A simple calculation gives the expected thermal spectrum
\be
\label{div}
\langle N(\omega)\rangle \; =
{A \over e^{2\pi\omega}-1}  - B
\ee
with
\be
\label{a}
 A = \int\limits_\omega^{\infty}
{d\xi\over \xi} \, \qquad \qquad B =  \int\limits_\omega^{\infty} {d\xi\over \xi}\,
{1\over e^{2\pi\xi} -1}
\, \nonumber .
\ee
The integral $A$ in (\ref{integral}) as it stands diverges; 
this divergence is a consequence of our idealized representation
of the incoming matter pulse. In the following section we will repeat the
corresponding calculation in the matrix model, while taking into
account that the incoming matter pulse contains a finite total energy $M$.

\newsubsection{Hawking Effect and Backreaction in the Matrix Model}

In this section, we will consider an incoming high energy pulse,
represented by a coherent state of the form
\be
\label{vefnew}
|\, {\it in}\, \rangle \, = \, V_h \; | \, 0 \, \rangle \, , \qquad \qquad 
V_h = \exp\Bigl(\int\!\! d u\, h(u) \partial_u C(u)\Bigr) 
\ee
with $h(u)$ some sharply localized function. We will find that the resulting outgoing state
will contain a thermal component, with the correct temperature
to be identified with Hawking radiation. The total energy contained in the
radiation is rather small, however: it scales as $\log M$ for a total 
incoming energy $M$.

We will be focusing on the high energy regime, in which, from the matrix model
perspective, the incoming signal consists of eigenvalues with energy $\omega >\!\!> \mu$.  
For simplicity, we will therefore set $\mu = 0$ in this section.

We start with a simple observation. Given the form (\ref{vefnew}) of the incoming
state, it immediately follows that in terms of the fermionic matrix model language, the final 
state will take the form $\exp {\cal B} %_h(\psi^\dagger,\psi)
 |\, 0 \, \rangle$ with ${\cal B}$ %_h(\psi^\dagger,\psi)$ 
some bilinear operator in terms of the fermions. The final state is therefore
uniquely specified once we know the two-point function of the outgoing fermions.
A special two-point function is the fermion number density as a function of frequency.
Let us define the 2$\times$2 matrix of number densities
\ba
N_{\alpha\beta}(\omega) & = & b^\dagger_{\alpha}(\omega)\, b_{\beta}(\omega)
\ea
where $\alpha$ and $\beta$ refer to the left or right half, see Eqn (\ref{expan}).  
The $in$ and $out$ number operators are related via the single fermion S-matrix via
\ba
N^{out}_{\alpha\beta}(\omega) = S^\dagger_{\alpha\gamma} N^{in}_{\gamma\delta}(\omega) S_{\delta\beta}
\ea
Using the explicit expression (\ref{smatrix}) for the S-matrix, we find
\ba
N^{out}_{++}(\omega) & = &  \frac{1}{1+e^{2\pi \omega}} N_{++}^{in}(\omega) + 
\frac{1}{1+e^{-2\pi\omega}} N_{--}^{in}(\omega) - i \frac{N_{-+}^{in}(\omega)
- N_{+-}^{in}(\omega)}{e^{\pi\omega} + e^{-\pi\omega}} \nonumber \\[-1mm]
\\[-1mm]
%N^{out}_{--}(\omega) & = & 
%\frac{1}{1+e^{2\pi \omega}} N_{--}^{in}(\omega) + \frac{1}{1+e^{-2\pi\omega}} N_{++} - i \frac{N_{+-}
%- N_{-+}}{e^{\pi\omega} + e^{-\pi\omega}}
%\\
N^{out}_{+-}(\omega) & = & \frac{1}{1+e^{2\pi \omega}} N_{+-}^{in}(\omega) + \frac{1}{1+e^{-2\pi\omega}} 
N_{-+}^{in}(\omega) - i \frac{N_{--}^{in}(\omega)
- N_{++}^{in}(\omega)}{e^{\pi\omega} + e^{-\pi\omega}} \nonumber %\\
%N^{out}_{-+}(\omega) & = & \frac{1}{1+e^{2\pi \omega}} N_{-+} + \frac{1}{1+e^{-2\pi\omega}} %N_{+-} - i \frac{N_{++}
%- N_{--}}{e^{\pi\omega} + e^{-\pi\omega}}
\ea
This equation is already quite instructive: it shows that, if we would consider an incoming excitation
localized on only one side of the well, so that only $N_{++}^{in}(\omega)$ is 
non-vanishing, the reflected part of the out-state has a number density given by
\be
\label{hoho}
N^{out}_{++}(\omega)= \frac{1}{1+e^{2\pi\omega}}N^{in}_{++}(\omega).
\ee
Therefore if the incoming fermion number density $N^{in}_{++}(\omega)$ would be constant over some sufficiently
large frequency range, the reflected part of the outgoing state would be thermal. In a moment we will 
show that, for a generic incoming localized matter pulse, $N^{in}_{++}(\omega)$ will indeed be constant
over a large range of frequencies.

General incoming signals will have support on both sides, so that other number densities besides $N_{++}^{in}$
will be non-vanishing. This spoils the thermal form of the outgoing fermion number density. Our interpretation
of the above formulas, however, is that the total scattering matrix naturally splits up into different 
S-matrix channels, each of which describes a different geometrical contribution to the total scattering 
process. Our main observation will be that the S-matrix channel associated with reflection off the matrix
model potential describes the part of the outgoing state produced via the Hawking effect. Indeed, the
thermal prefactor in  (\ref{hoho}) has the correct temperature to support this interpretation.

Since we have set $\mu=0$, reflection off the potential is in fact a subdominant effect for large
energies. In the regime of interest, the deformations of the Fermi sea extend far above the top
of the potential, and as a result, most of the incoming wave will be transmitted to the other side.
Via the matrix model/target space dictionary, this transmitted wave describes matter that is 
reflected back from the strongly curved target space region.  We will further comment on the 
interpretation of the transmitted signal in the next subsection.

To proceed, let us compute the incoming fermion number density associated with a given coherent state
of the form as given in (\ref{vef}). In fermionized form, it is given by
\be
V_f |\, 0\, \rangle = \exp\left[i \int\limits_{-\infty}^\infty\! du\, f(u) \psi^\dagger(u) \psi(u) \right] |\, 0\, \rangle.
\ee
$V_f$ can be thought of as an operator that implements a specific Bogolubov transformation on the
fermions. It satisfies the following exchange algebra with the fermion field
\be
\psi(u)\, V_f \, =\, V_f\, \psi(u)\, e^{-i f(u)}.
\ee
%Here we see the physical meaning of $f(x)$ -- since bosonization rules tell us that 
%$\psi(x) \sim e^{i \varphi(x)}$,
%$V_f$ adds the distribution $v(x)$ to $\varphi(x)$.   
Let us write $\Psi(u) = e^{-if(u)} \psi(u)$, and let us denote the frequency modes of $\Psi(u)$
by  $B_\omega$ and $B^\dagger_\omega$. We thus find for the incoming fermion number density
\be
\label{nino}
N^{in}(\omega) = \langle \, 0 \, | V_f \, b^\dagger_\omega b_\omega \, V_f\, | \, 0 \, \rangle 
= \langle\, 0\, | B^\dagger_\omega B_\omega |\, 0\,
\rangle = \int\limits_0^\infty {d\xi\over 2\pi} \, |\beta_{\omega\xi}|^2
\ee
where and $\beta_{\omega \xi}$ is the Bogolubov
coefficient in the relation $B_\omega = \alpha_{\omega \xi} b_\xi + \beta_{\omega\xi}
b^\dagger_\xi$ between the $B$ and $b$ modes. It given by the integral
\be
\beta_{\omega \xi} = \int\limits_{-\infty}^\infty \! du\, e^{i (\omega + \xi) u} e^{-i f(u)}
\ee
Computing this integral yields the incoming number distribution for any profile $f(u)$.
%Natsuume and Polchinski argued as much analytically in \cite{joepn}, and we have
%confirmed their arguments numerically.  The basic form of the integral is
%\be
%\varphi(x) \approx A\left[\frac{e^{(x - x_0)}}{1 + e^{(x-x_0)^3}} + cos(x) e^{-(x-x_0)^2/l^2}\right]
%\ee
%We can take this function to be our $v(x)$.
For our purpose, it is sufficient to evaluate it using  the stationary phase approximation. 
For each stationary point  $\bar{u}$, satisfying $f'(\bar{u}) = \xi$, one obtains a 
contribution $\sqrt\frac{2\pi}{f''(\bar{u})}$.
Plugging this back into (\ref{nino}) we find
\be
\label{nina}
N^{in}(\omega) \simeq \, \int \! \! \frac{d\xi}{f''(\bar{u})}
 \, \simeq\, \int \! d\xi {d\bar{u}\over d\xi}
\; \, %{v''(\bar{x})} \frac{d\xi'}{d\bar{x}} =
\simeq \; \bar{u}_{max}(\omega) - \bar{u}_{min}(\omega)
\ee
%where we used that ${d\xi\over d\bar{u}} = f''(\bar{u})$ and 
where $\bar{u}_{max}(\omega)$
and $\bar{u}_{min}(\omega)$ denote the maximum and minimum of the two solutions
to $f'(\bar{u}) = \omega$ (see Fig 2). As seen from the figure, this result has a clear
semi-classical interpretation: it tells us that the spectral fermion number density 
$N(\omega)d\omega$ is given by the phase space volume of the strip between $\omega$ and 
$\omega+d\omega$ that is filled by the perturbed Fermi sea.

\figuur{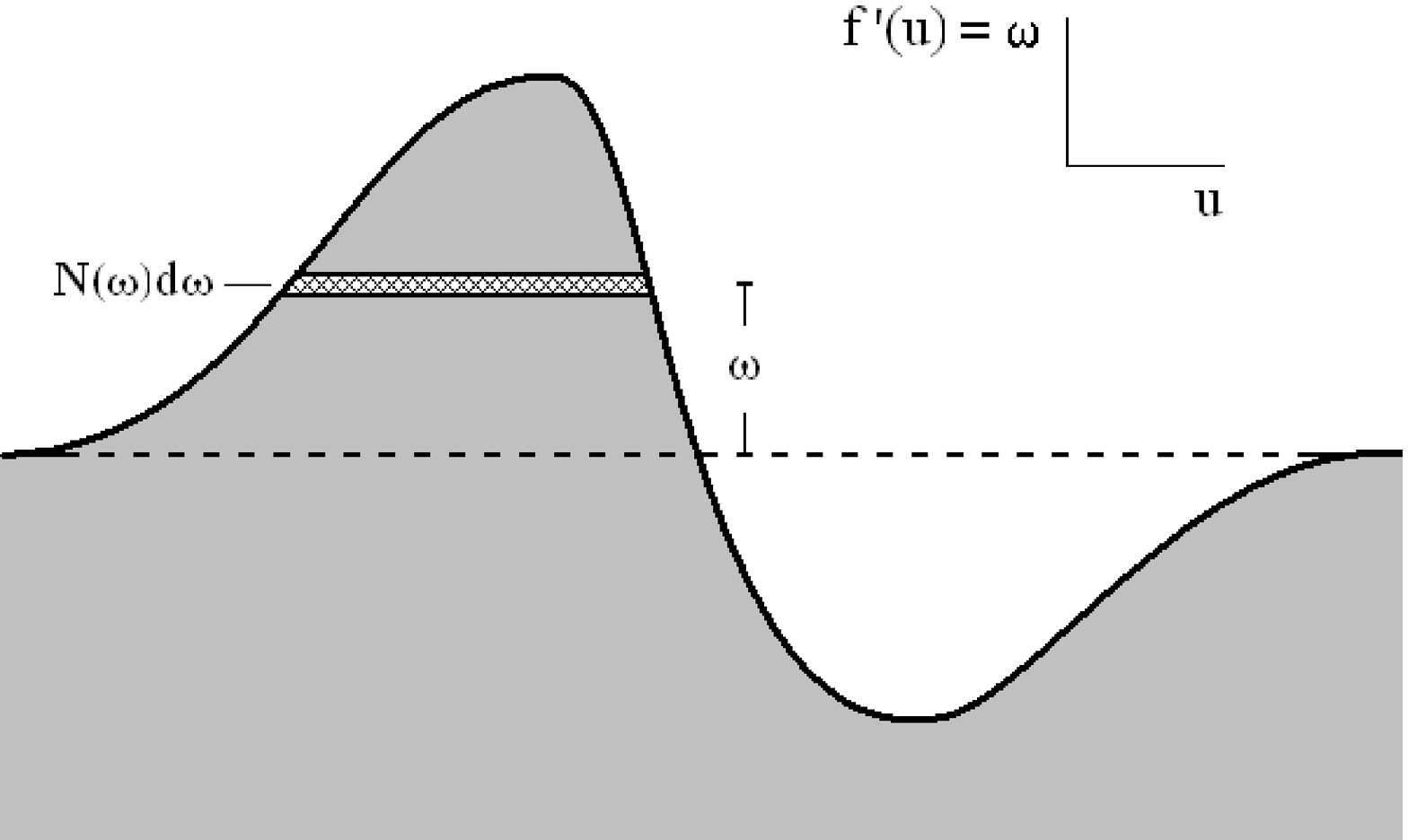}{8cm}{ Fig 2. The spectral fermion number density $N(\omega) d\omega$ is given 
by the phase space volume of the strip between $\omega$ and $\omega+d\omega$ that is filled 
by the perturbed Fermi sea.}

Next we need to determine the approximate shape of the Fermi sea corresponding to a localized
matter pulse in the target space. 
The profile $f(u)$ of the coherent state (\ref{vef}) of the matrix model scalar field $\varphi$ 
is related to the target space profile $h(u)$ in (\ref{vefnew}) via the inverse leg pole transform
\be 
\label{lepo}
f(u)  = {\Gamma({1\over 2}(1-\, \partial_u)) \over \Gamma({1\over 2}(1+ \partial_u) )} \;
\, h(u). 
\ee 
Since $h(u)$ is assumed to be sharply localized, we may approximate it by 
\be
h(u)\, \simeq\, H_0\, \delta(u-u_0), 
\ee
where the amplitude $H_0$ is assumed to be large. The properties of the leg pole transform
(\ref{lepo}) are discussed in some detail in \cite{joepn}. For our purpose, it will be 
sufficient to know the leading order behavior of $f(u)$ at early times relative 
to $u_0$. It turns out that in this regime, the non-local leg-pole transform is dominated by the 
pole in the function $\Gamma({1\over 2}(1-\, \partial_u))$ at $\partial_u =1$. So at
time $u$ sufficiently early 
compared to $u_0$, we can approximate
\be 
\label{rrr}
\qquad \qquad \qquad f(u) %\simeq\; h_0 (1- \partial_u)^{-1} \delta(u-u_0)\; 
\simeq\;  H_0\; e^{u-u_0}  \qquad \qquad u_0 - u >\!\!> 1.
\ee 
The contribution from the higher order poles lead to higher powers of $e^{u-u_0}$, and are therefore 
subdominant for $u_0\! -u>\!\!> 1$. As a related point, we remark that if, instead of the axion
field $C(u)$, we would consider an incoming tachyon matter wave, the corresponding leg pole transform 
would give $f(u)$ decaying as $e^{2(u-u_0)}$ -- tachyon matter waves in this respect are subdominant
to axion matter waves.

Given the result (\ref{rrr}), we can now estimate the incoming fermion number 
density $N^{in}(\omega)$ for frequencies $\omega$ much smaller than $H_0$, via (\ref{nina}).
From $f'(\bar{u}_{min}) = \omega$ and (\ref{rrr}) we find $\bar{u}_{min} 
\simeq u_0 - \log (H_0/\omega)$. Setting $\bar{u}_{max} \simeq u_0$ we obtain
\be
N^{in}(\omega) \simeq \log(H_0/\omega). 
%\simeq \log M - \log \omega
\ee
Combined with (\ref{hoho}), this tells us that the reflected out-signal indeed has an (approximately)
thermal spectrum of the form\footnote{
Since the amplitude $H_0$ is assumed to be very large 
compared to the typical Hawking frequency $\omega$, we dropped the subleading $\log \omega$-term.}
\be
\label{fin}
N^{out}_{refl}(\omega) \, \simeq \, {\log H_0 \over e^{2\pi \omega} +1}\, .
\ee 
It is interesting to compare this result with the divergent answer (\ref{div})-(\ref{a}) 
obtained earlier, via an idealized calculation in the low energy effective field 
theory. After bosonizing (\ref{fin}), we can directly compare the two. We note
that the matrix model result essentially amounts to a UV regulation of the frequency 
integral $A$ in (\ref{a}), cutting it off at at a frequency $\xi_{max} = H_0$
equal to the maximal energy carried by any of the matrix model eigenvalues.
Conversely, within the matrix model, one can formally recover the exact result
(\ref{bcoef}) for the Bogolubov coefficients by taking the limit $H_0\to \infty$,
$u_0 \to \infty$, while keeping $H_0 e^{-u_0}$ fixed.

Based on the correspondence with the semi-classical target space physics, 
we would like to make the following identification between the amplitude $H_0$,
the instant $u_0$ and the total incoming Kruskal momentum $p_+$ of the incoming matter pulse
\be
\label{wellwell}
H_0\, e^{-u_0}\, \simeq\, p_+/\lambda
\ee
where $\lambda$ is the coefficient of $e^{-2\phi}\sqrt{-g}$ in the dilaton gravity 
Lagrangian (\ref{act}). 
Combined with (\ref{rrr}), this identification tells us that relative to 
operators defined at early times, the operator $V_f$ behaves as
\be
\label{wish}
V_f \, \sim \; e^{{i\over \lambda} p_+ Q^+ } 
\ee
with 
\be
Q^+ = \int\limits_{-\infty}^{\infty}
\!\! du \, e^u
\, \partial_u \varphi_{in}
= \int\limits_0^\infty \! dx :\! \psi^\dagger_{in}(x)  {x} \, \psi_{in}(x)\! :
\ee 
In terms of the $out$-modes,
\be
Q^+ \, =\, 
\int\limits_0^\infty \! dq  :\!\psi^\dagger_{out}(q) \, q \, \psi_{out}(q)\!: 
\ee
with $\psi_{out}(q)$ the outgoing fermionic mode with given Kruskal momentum $q$. 
We see that $Q^+$ represents the total outgoing Kruskal momentum, restricted to 
the subsector $q>0$.
Consequently, the operator $V_f$ that creates the incoming matter pulse has the 
following effect on the out-modes (with $q>0$)
\be
\label{shiftk}
\psi_{out}(q)\, V_f\, =\, V_f \, \psi_{out}(q)\, e^{{i\over \lambda} p_+ q}
\ee
This relation can be recognized as the momentum space version of the target space 
exchange algebra (\ref{shft}): it tells us that, also in the matrix model, an
incoming matter pulse with Kruskal momentum $p_+$ results in a coordinate
shift identical to (\ref{shifto}) on (part of) the outgoing modes. We believe this 
correspondence provides an important hint for how the  
gravitational target space dynamics is encoded in the matrix model, 
as well as concrete support for the interpretation of the reflected wave
as Hawking radiation.

The proposed linear relation (\ref{wellwell}) between the amplitude $H_0$ and the 
Kruskal momentum
is directly motivated by the correspondence between (\ref{shiftk}) and (\ref{shft}).
It has actually a slightly unexpected form, since energy-momentum is typically proportional to the
{\it square} of the amplitude. Suppose, however, that the vacuum state satisfies 
\be
\label{lola}
\int\limits_0^\infty \! dx  \! :\! \psi^\dagger (x) \psi(x)\! : |\, 0 \, \rangle\, =\, 
\lambda \, | \, 0 \, \rangle
\ee
From Eqns (\ref{cancan}), (\ref{lll}) and (\ref{wish}), we can then deduce that the 
$in$ state $V_f | \, 0 \, \rangle$ is indeed an (approximate) eigenstate of the total 
Kruskal momentum with eigenvalue $p_+$.  In bosonized language, (\ref{lola}) translates to 
a finite, non-zero vacuum value for the zero-mode $\int \partial \varphi$ of the 
matrix model scalar field. 

The relation (\ref{lola}) looks rather 
subtle, since typically one would define the normal ordered expression such
that the right-hand side vanishes. Our proposed interpretation, however, is as follows. 
In order to establish a correspondence with target space physics, one must find
the proper identification of the target space quantities such as the total
energy and total incoming and outgoing Kruskal momenta. These quantities are 
defined as normal ordered expressions as in (\ref{www}) and (\ref{kmom}).
The meaning of Eqn (\ref{lola}) is that it specifies the normal 
ordering prescription one needs to use in defining the space-time charges.

We end with a comment concerning the tachyon field.
Given that, relative to the axion field, a localized tachyon matter wave gives rise 
to a subdominant Fermi sea profile for early times $u$, we must conclude that a 
collapsing shell of tachyon matter leads to a subdominant, or perhaps even 
absent, gravitational Hawking effect. This is somewhat mysterious, since it implies
that the gravitational backreaction due to a tachyon matter pulse is much less
pronounced than that of an axion matter pulse. Presumably, this difference
arises due to strong self-interactions of the tachyon.

\newsubsection{Fate of the Matter Pulse}

We have identified the Hawking radiation in the matrix model as the signal that reflects off
the inverted harmonic potential. In the high energy regime, where the deformations of the Fermi 
sea extend far above the top of the potential, most of the signal simply propagates through to the
other side. From the target space perspective, this transmitted wave describes matter that is 
reflected back via some non-linear dynamics that takes place in the strongly curved region.
Unlike most higher dimensional black holes, the surface gravity of the black hole
in 2-d string theory is independent of its mass or size, and always of order the string scale.
Stringy effects can therefore kick in, and lead to drastic modifications of the low energy
effective dynamics before the infalling matter has a chance to retreat behind the horizon. 
Our present calculation indicates that such a non-linear modification indeed takes place,
and results in a complete reflection of all infalling matter.

\figuur{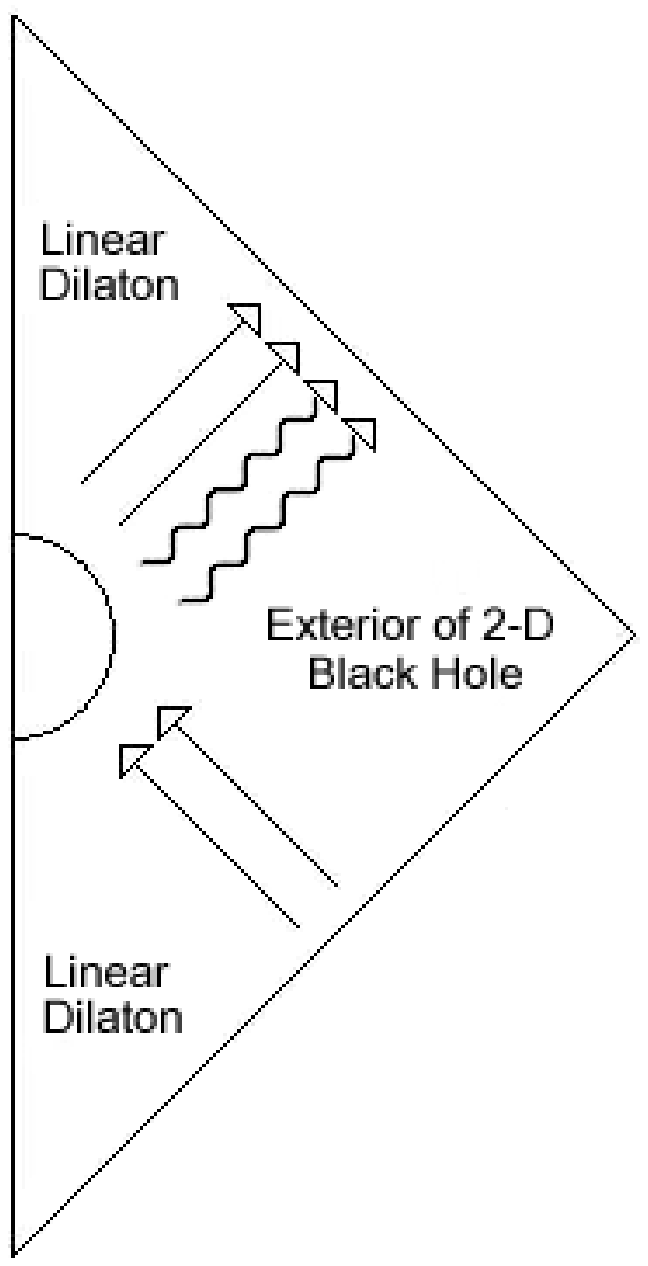}{13cm}{Fig 3. The space-time interpretation of the high energy scattering in
the matrix model. The straight arrows represent infalling matter that gets reflected back 
via a non-linear process in the strongly curved region. %This reflection takes place before any 
%black hole horizon can be formed. 
The wavy arrows represent geometric Hawking radiation that 
appears a short time (of order $\log M$) before most of the matter gets returned. }

Given our simple geometric characterization of the S-matrix as Fourier transformation in Kruskal
coordinates, it is easy to determine the rough location where the reflection takes place.
Suppose we consider a incoming matrix model eigenvalue with a wave function of the form
\be
\psi_{q,p}(x) = e^{-{1\over 2} (x-q)^2} e^{ip x}
\ee
This eigenvalue has an approximate Kruskal position equal to $x \simeq q$ and incoming
Kruskal momentum equal to $p$.
Its energy is approximately $\omega \simeq p\, q$. Now, the S-matrix simply exchanges $p$ and $q$:
\be
\psi_{q,p} = S\, \psi_{p,q}\, .
\ee
The effective location of reflection is therefore given by $x^+x^- \simeq \omega$. 
As seen from (\ref{rrr}), the most energetic matrix model eigenvalue has an energy 
proportional to $H_0$, which scales as $\sqrt{M}$. 
A good estimate for the outgoing time $v_{refl}$, when most of the matter wave 
gets returned, is therefore $v_{refl} \simeq\, u_0 - {1\over 2} \log M$. This 
should be compared with the moment when the Hawking radiation turns on,
which is roughly at $v_{rad} \simeq \, u_0 - \log M$. According to these estimates,
the Hawking radiation precedes the return of the large matter pulse by a short time 
interval of order ${1\over 2} \log M$ (see Fig 2).

\newsubsection{Conclusion}

We have proposed a dictionary between the matrix model and two-dimensional
string theory in the high energy regime where the semi-classical gravitational backreaction
of the incoming matter wave is expected to give rise to black hole radiance. We have found that
the Hawking effect indeed takes place, but that the Hawking radiation carries off only a 
relatively small portion of the total energy $M$, proportional to $\log H_0$ where $H_0$ is the 
amplitude of the incoming matter wave. Given that $M \sim H_0^2$ and that the Hawking flux is of
order one in string units, this tells us that the radiation is produced for a short amount of time 
of order $\log M$.

Although our finding that high energy incoming matter fails to produce a long-lived black hole 
counts as somewhat of a disappointment, our paper still supports some positive conclusions.
We have found, within the matrix model, a clear signal of semi-classical target space physics.
As such, our results represent one of the first time-dependent tests of the correspondence principle 
between a holographic theory and its gravitational dual.

\bigskip

\bigskip
\bigskip

\noindent
{\sc Acknowledgements}

\medskip

\noindent
We would like to thank S. Das, O. DeWolfe, S. Giddings, J. Maldacena, E. Martinec, J.
McGreevy, A. Strominger, L. Thorlacius, and S. Wadia for helpful discussions.  HV's work
was supported in part by NSF grant No. 0243680.  JF's work was funded in
part by the National Science Foundation Graduate Research Fellowship
Program.  Any opinions, findings, and conclusions expressed in this
material are those of the authors and do not necessarily reflect the views
of the National Science Foundation.

\newpage

\bibliographystyle{utphys}
\bibliography{smatrix}

\end{document}